\documentclass[fleqn,usenatbib]{mnras}

\usepackage{newtxtext,newtxmath}

\usepackage[T1]{fontenc}

\DeclareRobustCommand{\VAN}[3]{#2}
\let\VANthebibliography\thebibliography
\def\thebibliography{\DeclareRobustCommand{\VAN}[3]{##3}\VANthebibliography}

\usepackage{graphicx}	
\usepackage{amsmath}	
\usepackage[normalem]{ulem}

\title[Accretion bottleneck]{Accretion bottleneck in protoplanetary discs: the role of the stellar spin}

\author[Longarini \& Clarke]{
Cristiano Longarini$^{1}$\thanks{E-mail: \url{cl2000@cam.ac.uk}} and
Cathie J. Clarke$^{1}$
\\
$^{1}$Institute of Astronomy, University of Cambridge, Madingley Rd, CB30HA, Cambridge, UK
}

\date{Accepted XXX. Received YYY; in original form ZZZ}

\pubyear{\the\year{}}

\begin{document}
\label{firstpage}
\pagerange{\pageref{firstpage}--\pageref{lastpage}}
\maketitle

\begin{abstract}
We investigate angular momentum transport and accretion properties in a sample of protoplanetary discs with dynamical measurements of stellar masses, disc masses, and scale radii. From these data we infer effective $\alpha$-viscosities, finding a remarkably broad range spanning over three orders of magnitude. This spread correlates with the stellar rotation period: systems with shorter periods exhibit significantly lower accretion rates, suggesting that they are undergoing at least temporary episodes of accretion bottleneck. We interpret this behaviour within the framework of magnetospheric accretion models, where the transition between steady accretion and the propeller regime is set by the relative locations of the co-rotation and magnetospheric radii. Our results indicate that stellar spin is a key parameter in regulating mass transfer from the disc to the star, and provide new evidence that the observed dispersion in $\alpha$ reflects transitions between distinct accretion states rather than differences in global disc properties.

\end{abstract}




\section{Introduction}
The physical mechanism responsible for angular momentum transport in protoplanetary discs remains uncertain. In the earliest phases of disc evolution, particularly in young and massive systems, gravitational instabilities are expected to play a major role \citep{kratter16}. At later stages, however, the picture is less clear. Magneto-hydrodynamical (MHD) processes such as the magneto-rotational instability \citep{balbus91} or magneto-hydrodynamic winds \citep{tabone22} have been considered a promising mechanism, but the former requires a degree of ionisation in the disc material that is poorly constrained both observationally and theoretically, and the latter that there is a net magnetic field threading the disc.

A common way to work around this uncertainty is to adopt a viscous prescription, in which angular momentum transport is parametrised through the so-called $\alpha$-viscosity \citep{shakura73}
\begin{equation}
    \nu = \alpha c_s H,
\end{equation}
where $c_s$ is the sound speed and $H$ the disc scale height. The dimensionless parameter $\alpha$ describes the efficiency of angular momentum transport, regardless of the underlying physical process. Within this framework, classical viscous theory links disc properties, such as mass, size, and temperature, with the accretion rate onto the central star. Measuring accretion rates therefore offers, in principle, a way to constrain the instantaneous value of $\alpha$. We point out that here we are assuming that stellar accretion, which happens in the <1 au region of the disc, is equal to the outer disc accretion, measured at the scale radius. Several studies have pursued this approach \citep{andrews09,andrews10,rafikov17,ansdell18,vandermarel21}, consistently finding that values $\alpha \gtrsim 10^{-4}$ are required to explain the observed accretion rates. These analyses, however, relied on dust-based estimates of disc masses and scale radii, either inferred from the continuum emission or assumed to scale with the flux of the $^{12}$CO line.

In this work, we make use of high-resolution ALMA observations \citep{oberg21,teague25} that provide dynamical measurements of stellar masses, disc masses, and disc sizes \citep{martire24,Longarini25}. With these improved constraints, we are able to revisit the $\alpha$-viscosity problem and examine its implications for accretion processes in protoplanetary discs. This Letter is structured as follows. In Section~\ref{method} we describe the method used to infer $\alpha$-viscosities from dynamical disc properties and present the resulting distribution across our sample. In Section~\ref{discussion} we explore possible origins of the observed spread, highlighting the emerging correlation with stellar rotation period and its interpretation within magnetospheric accretion models. Finally, in Section~\ref{conclusions} we summarise our main findings and discuss their implications for the regulation of mass transfer in young stellar systems.

\section{Method}\label{method}

\begin{table*}\renewcommand{\arraystretch}{1.25}
    \centering
    \caption{Sample's properties: star mass, disc mass, scale radius, aspect ratio at the scale radius, accretion rate, $\alpha$-viscosity and $r$ parameter.\newline Stellar rotation period references: AS209, AA Tau \citet{artemenko12}; DM Tau, J1842 and J1852 \citet{pittman25};  GM Aur \citet{percy2010}; IM Lup \citet{siwack16} ; J1615 \citet{deboer16}; LkCa15 \citet{alencar18}; PDS66 \citet{batalha98}; SY Cha \citet{schaefer83}; V4046 \citet{stempels04}}
    \begin{tabular}{lccccccc}
        Source & $M_\star$ [M$_\odot$] & $M_d$ [M$_\odot$] & $R_c$ [au] & $H_c/R_c$ &  $\log_{10}\dot{M}_\star$ [M$_\odot$/yr] & $\alpha$ & $P_{\rm rot}$ [days] \\
        \hline
        AS 209  & $1.311 \pm 0.001$ & $<0.065$ & $126 \pm 2$ & $0.097$ & $-7.3 \pm 0.35$ & $>1.11\times10^{-2}$ & $8.60\pm0.43$ \\
        GM Aur  & $1.128 \pm 0.002$ & $0.118 \pm 0.002$ & $96 \pm 1$ & $0.083$ & $-8.1 \pm 0.35$  & $9.32^{+11.1}_{-5.16}\times10^{-4}$ & $6.10\pm0.30$ \\
        HD 163296  & $1.948 \pm 0.002$ & $0.134 \pm 0.001$ & $91 \pm 1$ & $0.068$ & $-7.4 \pm 0.35$ & $4.31^{+5.35}_{-2.38}\times10^{-3}$ &  ... \\
        IM Lup  & $1.194 \pm 0.002$ & $0.106 \pm 0.002$ & $115 \pm 1$ & $0.099$ & $-7.9 \pm 0.35$  & $1.48^{+1.18}_{-0.82}\times10^{-3}$ & $7.24\pm 0.03$ \\
        MWC 480  & $2.027 \pm 0.002$ & $0.150 \pm 0.002$ & $128 \pm 1$ & $0.081$ & $-6.9 \pm 0.35$ & $1.41^{+1.17}_{-0.78}\times10^{-2}$ & ...  \\
        AA Tau  & $0.624^{+0.033}_{-0.035}$ & $0.155^{+0.036}_{-0.036}$ & $156^{+106}_{-41}$ & $0.195$ & $-7.3 \pm 0.35$  & $6.54^{+8.11}_{-3.62}\times10^{-3}$ & $8.19\pm0.41$ \\
        DM Tau  & $0.468^{+0.014}_{-0.015}$ & $0.057^{+0.019}_{-0.020}$ & $240^{+42}_{-27}$ & $0.236$ & $-8.2 \pm 0.35$ & $2.19^{+2.72}_{-1.21}\times10^{-3}$ & $7.87\pm1$ \\
        HD 34282  & $1.520^{+0.025}_{-0.031}$ & $0.143^{+0.045}_{-0.041}$ & $370^{+109}_{-78}$ & $0.219$ & $-7.7 \pm 0.35$ & $3.86^{+4.79}_{-2.14}\times10^{-3}$ & ... \\
        RXJ1615  & $1.105^{+0.011}_{-0.012}$ & $0.082^{+0.014}_{-0.014}$ & $167^{+20}_{-15}$  & $0.135$ & $-8.2 \pm 0.35$ & $1.20^{+1.48}_{-0.66}\times10^{-3}$ & $5.72\pm0.01$ \\
        RXJ1842 & $1.042^{+0.010}_{-0.011}$ & $0.078^{+0.013}_{-0.014}$ & $231^{+102}_{-50}$ & $0.179$ & $-8.8 \pm 0.35$ & $4.03^{+4.99}_{-2.23}\times10^{-4}$ & $2.54\pm0.1$ \\
        RXJ1852 & $1.022^{+0.021}_{-0.021}$ & $0.044^{+0.024}_{-0.032}$ & $87^{+69}_{-16}$  & $0.121$ & $-8.7 \pm 0.35$ & $3.62^{+4.49}_{-2.01}\times10^{-4}$ & $2.83\pm0.13$ \\
        LkCa15 & $1.118^{+0.013}_{-0.015}$ & $0.108^{+0.016}_{-0.016}$ & $150^{+12}_{-16}$  & $0.154$ & $-8.7 \pm 0.35$ & $1.74^{+2.16}_{-0.97}\times10^{-4}$ & $5.70\pm0.15$ \\
        PDS66 & $1.299^{+0.036}_{-0.101}$ & $0.038^{+0.099}_{-0.035}$ & $28^{+12}_{-5}$  & $0.060$ & $-9.2 \pm 0.35$ & $8.06^{+9.99}_{-4.46}\times10^{-5}$  & $5.75\pm0.03$ \\
        SY Cha & $0.812^{+0.037}_{-0.041}$ & $0.084^{+0.044}_{-0.043}$ & $112^{+21}_{-15}$  & $0.156$ & $-9.9 \pm 0.35$ & $1.66^{+2.05}_{-0.92}\times10^{-5}$  & $6.13\pm0.03$ \\
        V4046 Sgr & $1.777^{+0.005}_{-0.006}$ & $0.058^{+0.006}_{-0.006}$ & $99^{+5}_{-5}$  & $0.091$ & $-9.3 \pm 0.35$ & $1.11^{+1.38}_{-0.61}\times10^{-4}$ &  $2.42\pm0.01$ \\
    \end{tabular}
    \label{tab:properties}
\end{table*}

\begin{figure*}
    \centering
    \includegraphics[width=\linewidth]{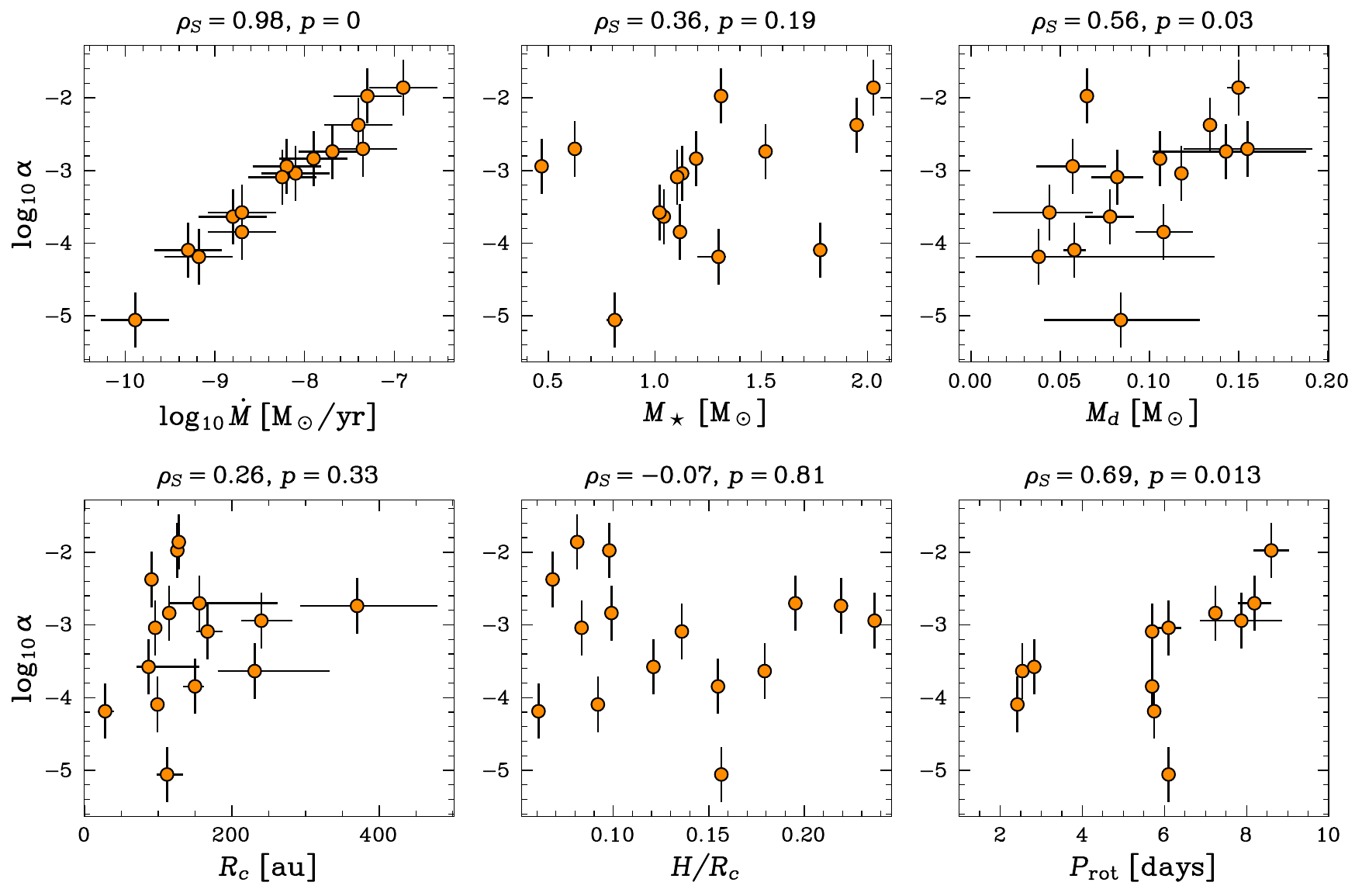}
    \caption{Scatter plots of $\alpha$ against the sample's properties, stellar accretion rate, star mass, disc mass, scale radius, aspect ratio and stellar rotation period (see Table \ref{tab:properties} for reference). {On the top of each panel the value of the Spearman rank coefficient $\rho_S$ and $p$-value are shown.}}
    \label{fig:correlations}
\end{figure*}

\cite{martire24} and \cite{Longarini25} estimated the dynamical masses and sizes of MAPS and exoALMA discs by modelling the rotation curves of $^{12}$CO and $^{13}$CO. All the sources have measured accretion rates onto the central object $\dot{M}_\star$ (references can be found in \citealt{Longarini25}), and this made possible to compute $\alpha-$viscosities, using (e.g. \citealt{hartmann88})
\begin{equation}\label{alpha}
    \alpha = \frac{2\dot{M}_\star}{3M_d\Omega_c}\left(\frac{H}{R}\right)_c^{-2},
\end{equation}
where $\Omega = \sqrt{GM/R^3}$ is the Keplerian angular frequency, $H/R$ is the aspect ratio of the disc and the subscript $c$ indicates that the quantity is evaluated at the scale radius $R_c$ (see eq. \ref{selfsimilar} below). The aspect ratio $H/R$ is related to the disc temperature $T$ through hydrostatic equilibrium. To constrain this quantity, optically thick molecular tracers such as $^{12}$CO and $^{13}$CO can be used, as done in \cite{law21b, galloway25}. We underline that in \cite{martire24} and \cite{Longarini25} the underlying hypothesis is that the surface density profile follows a self-similar solution
\begin{equation}\label{selfsimilar}
    \Sigma = \frac{M_d}{2\pi R_c^2}\left(\frac{R}{R_c}\right)^{-1}\exp\left[-\frac{R}{R_c}\right].
\end{equation}
The values of $M_\star$, $M_d$, $R_c$, $\dot{M}_\star$ and $\alpha$ are shown in Table \ref{tab:properties}. {The possible impact of source inclination is discussed in \citet{Izquierdo25,Longarini25}. To ensure robust dynamical estimates of these quantities, we exclude sources with either very high or very low inclinations.} Interestingly, as noted by \cite{Longarini25}, the range of $\alpha$ is very broad, spanning three orders of magnitude from $10^{-5}$ to $10^{-2}$. 

Why is the inferred value of $\alpha$ so widely spread across the sample? Two broad interpretations can be considered. Either different sources have intrinsically different values of $\alpha$, reflecting distinct structural properties, or the observed spread is the consequence of temporal variability, with each system being caught at a different evolutionary stage. If $\alpha$ correlates with some disc property, the former interpretation would be favoured. However the implied disc lifetimes $(M_d/\dot{M}_\star)$ would become unrealistically long in the case of systems with low  $\alpha$ if $\alpha$ was truly temporally constant. A more plausible picture is therefore that both effects play a role.

We can formalise this in three possible scenarios:
\begin{itemize}
    \item $\alpha$ depends on a system property that remains invariant over its lifetime (e.g. the stellar mass), such that each system has a fixed value;
    \item $\alpha$ depends on a property that evolves during the disc lifetime, and therefore varies with time;
    \item $\alpha$ is uncorrelated with disc properties, and its variability instead reflects episodic accretion.
\end{itemize}
Figure \ref{fig:correlations} shows $\alpha$ as a function of stellar accretion rate, stellar mass, disc mass, scale radius, aspect ratio, and stellar rotation period for the sources in our sample (see Table \ref{tab:properties} for reference). Before discussing the three scenarios, it is important to note the clear correlation between $\alpha$ and the stellar accretion rate. This correlation arises by construction (see eq. \ref{alpha}), given the limited dynamic range of the other disc parameters. Consequently, expressing results in terms of accretion rate or in terms of $\alpha$ is effectively equivalent.

{To quantify the correlation, we evaluate the Spearman rank correlation coefficient, $\rho_{\rm S}$, which is defined as
\begin{equation}
    \rho_S = 1 - \frac{6\sum_i d_i^2}{n(n^2-1)},
\end{equation}
where $d_i$ is the difference between the ranks of the $i$-th pair and $n$ is the number of data points. The Spearman coefficient ranges from $-1$ (perfect anticorrelation) to $+1$ (perfect correlation), with $\rho_{\rm S}=0$ indicating no correlation.}

\subsection{First scenario: invariant $\alpha$}
According to the first scenario, $\alpha$ depends on a system property that remains invariant over the disc lifetime, so that each system retains a fixed value. The only such invariant quantity is the stellar mass, yet Figure \ref{fig:correlations} (upper central panel) shows no correlation between $\alpha$ and $M_\star${, with $\rho_S=0.36$ and $p=0.19$}. This lack of correlation provides no support for the first hypothesis, and is also consistent with the lifetime argument discussed above. 

\subsection{Second scenario: evolving $\alpha$}
According to the second scenario, $\alpha$ depends on disc properties that vary over time, implying that $\alpha$ itself evolves. According to classical accretion theory, quantities such as disc mass, scale radius, and disc temperature do evolve. However, figure \ref{fig:correlations} shows that there are no clear correlations between $\alpha$ and such quantities. {The only disc property that suggests a possible correlation is the disc mass, with $\rho_S=0.56$ and $p=0.03$, which corresponds to a significance of roughly $2\sigma$.}

Interestingly, the bottom right panel of Figure \ref{fig:correlations} shows a correlation between $\alpha$ and the stellar rotation period $P_{\rm rot}$, which is available just for a subset of our sample (see Table \ref{tab:properties}). Systems with lower $\alpha$ appear to be preferentially associated with shorter stellar rotation periods, with a sharp transition at $\sim$6 days. In general, we expect the stellar rotation period to evolve with time \citep{armitage96,matt10,matt12,gallet19}, supporting the second scenario. In section \ref{prot_discussion} we will discuss the physical reason of such correlation.

We find $\rho_{\rm S} = 0.69$ with a corresponding $p$-value of $p = 0.013$, indicating that the correlation between $\alpha$ and $P_{\rm rot}$ is statistically significant. The p-value represents the probability of obtaining a correlation at least as strong as the observed one under the null hypothesis of no correlation; values below 0.05 are typically considered statistically significant. We also tested the correlation between $\dot{M}$ and $P_{\rm rot}$, obtaining $\rho_{\rm S} = 0.73$ and $p = 0.007$, which suggests an even stronger correlation between the stellar rotation period and the accretion rate.

\subsection{Third scenario: \textit{random} $\alpha$}
According to the third scenario, the spread in $\alpha$ arises from intrinsic variability in the accretion rate, with sources caught during different episodic accretion phases. In this picture, $\alpha$ would be uncorrelated with any stellar or disc property, and the observed dispersion would be entirely due to temporal fluctuations. However, the evidence for a correlation between $\alpha$ and stellar rotation period argues against this interpretation. While both $\alpha$ and $P_{\rm rot}$ may indeed vary over time within a given source, the existence of a systematic relationship between the two quantities indicates that source-to-source differences in $\alpha$ cannot be explained solely in terms of random temporal variability.

\section{Discussion}\label{discussion}
\begin{figure}
    \centering
    \includegraphics[width=\linewidth]{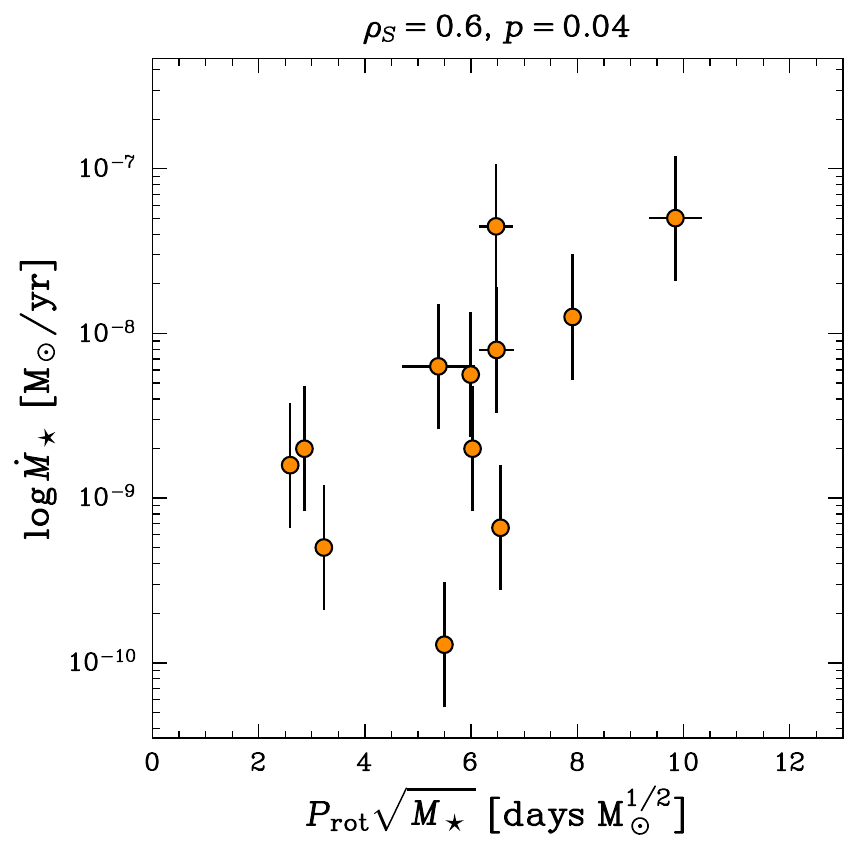}
    \caption{Accretion rate as a function of {$P_{\rm rot}\sqrt{M_\star}$} for the nine sources in our sample with rotation period measurements available in the literature.}
    \label{fig:mdot_prot}
\end{figure}

\subsection{Linking stellar spin and accretion rate}\label{prot_discussion}
Figure~\ref{fig:mdot_prot} shows the aforementioned correlation, but in the the {$\dot{M}_\star - P_{\rm rot}\sqrt{M_\star}$ space for the sub-sample of discs with available rotation measurements. We adopt $P_{\rm rot}\sqrt{M_\star}$ because it is directly linked to the co-rotation radius, the location in the disc where the angular velocity equals the stellar rotation rate. Notably, systems with lower accretion rates tend to show smaller values of $P_{\rm rot}\sqrt{M_\star}$.}

This behaviour is qualitatively consistent with the expectations from magnetospheric accretion models. In the standard picture of disc braking, accretion proceeds smoothly as long as the stellar magnetosphere truncates the disc inside the co-rotation radius (namely $R_M < R_{\rm co}$, where $R_M$ is the magnetospheric radius). For slowly rotating stars, the magnetosphere is well within co-rotation and the mass supply from the outer disc is efficiently channelled onto the star. When, however, the truncation radius approaches or exceeds the co-rotation radius (the so-called propeller regime), accretion is suppressed or becomes intermittent, since the magnetospheric interaction injects angular momentum into the disc material and inhibits mass inflow \citep[e.g.][]{romanova09,pantolmos20,ireland22}. Short-term simulations suggest three regimes depending on the ratio $R_{ M}/R_{\rm co}$: (i) $<0.43$ (steady accretion), (ii) $\in [0.43,1]$ (intermittent accretion), and (iii) $>1$ (propeller state, \citet{ireland22}). Around $P_{\rm rot}\sim 6$ days, the observed systems populate a wide range of $\alpha$ values (Fig. \ref{fig:correlations}), including very low values orders of magnitude below those inferred for the slowest rotators, consistent with systems switching between regimes (ii) and (iii). 

A persistent low-accretion state cannot represent a long-term steady solution, as it would imply unrealistically long disc lifetimes $M_d/\dot{M}_\star$. A more plausible explanation is that such states occur when the magnetospheric truncation radius lies very close to the co-rotation radius (i.e. the \textit{intermittent accretion state}). In this configuration, the system can switch between two regimes: a \textit{propeller state}, in which accretion onto the star is strongly suppressed, and a \textit{steady accretion state}, in which the stellar accretion rate reflects the mass supply from the outer disc. Even in the short-term calculations of \citet{ireland22}, where $P_{\rm rot}$ and $B_\star$ remain essentially fixed, large temporal fluctuations in the accretion rate are observed once the system enters propeller mode, albeit with peak accretion levels well below those achieved when $R_M \ll R_{\rm co}$. Such variability is in fact necessary in order to explain why these massive gas discs do not survive for $\sim$Gyr timescales. This provides a plausibility argument that the very low accretors in our sample do not remain permanently in this state; instead, accretion is likely to proceed in bursts once the stellar rotation period approaches $\sim 6$ days or below. {It is important to point out that the quantity that determines the accretion state is the corotation radius rather than the stellar rotation period, which also depends on the stellar mass, but that the range in stellar mass in our sample is small. }

One possible way to exit the propeller state requires the star to spin down, so that the magnetospheric radius moves inside the co-rotation radius. The stellar spin-down timescale therefore provides an important constraint on the regulation of accretion in young stars. Estimates by \citet{ireland22} suggest that the time required for the star to slow down by a factor of two is of order $\sim$1 Myr. In the propeller regime, however, this timescale depends mainly on the truncation radius, which is only weakly sensitive to the accretion rate (see right-hand side of Fig.~4 in \citealt{ireland22}).

For systems where $R_M$ and $R_{\rm co}$ are close to each other, modest modulations of either radius can be sufficient to switch between accretion and propeller regimes. While order-unity changes in stellar spin occur on Myr timescales, the much smaller adjustment in spin period required to trigger a transition can take place on a correspondingly shorter timescale. In addition, shorter-term variations in $R_M$ associated with stellar magnetic activity cycles \citep{clarke95} can also precipitate such changes, allowing episodic accretion even in systems nominally in the propeller state.

\begin{figure}
    \centering
    \includegraphics[width=\linewidth]{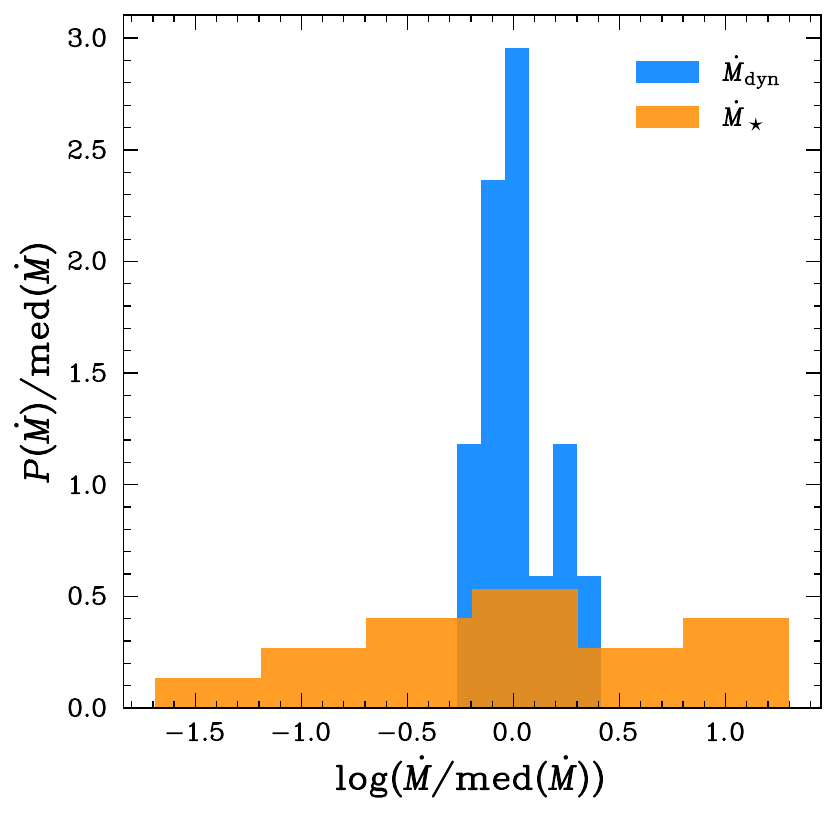}
    \caption{{Distribution of $\dot{M}_\star$ and $\dot{M}_{\rm dyn}$ normalised to their median value, showing that the spread in accretion rates is more than an order of magnitude larger than what would be expected due to the range of outer disc properties in the sample. }}
    \label{fig:spread}
\end{figure}

A useful comparison can be made with the results of \citet{venuti17}, who studied the relationship between stellar rotation period and accretion rate in the young cluster NGC~2264. In their sample, no clear correlation between $\dot{M}\star$ and $P{\rm rot}$ was detected, with the accretion rate exhibiting a large scatter at all rotation periods. This difference with respect to our findings may be explained by the much broader range of stellar and disc properties in their sample (e.g. stellar mass, disc mass, outer disc radius). In systems with such diverse outer disc properties, the mass flux feeding the inner disc varies substantially, leading to different truncation radii. As a result, the critical rotation period at which systems approach the propeller regime—and the associated decline in accretion rate—also exhibits a larger dispersion, which can wash out any underlying trend. By contrast, our sample consists of large, bright discs with relatively homogeneous outer disc properties and similar mass transport rates. {To demonstrate this homogeneity, we re-express equation \ref{alpha} as
\begin{equation}
    \alpha = \frac{\dot{M}_\star}{\dot{M}_{\rm dyn}},
\end{equation}
where $\dot{M}_{\rm dyn} = 3/2M_d\Omega_c(H/R)^2_c$ is the expected accretion rate in the outer disc  if $\alpha$ were locally equal to 1. Notably $\dot{M}_{\rm dyn}$  only depends on disc mass and the orbital frequency and aspect ratio in the outer disc. Figure \ref{fig:spread} shows the distribution of $\dot{M}_\star$ and $\dot{M}_{\rm dyn}$ normalised to their median value. The figure clearly shows that the spread in accretion rates onto the star is more that an order of magnitude larger than the expected range in accretion rates given the variation in outer disc properties contributing to $\dot{M}_{\rm dyn}$.}
Once this source of variability is controlled, an additional dependence of the accretion rate on stellar spin emerges, suggesting that stellar rotation plays a previously hidden role in regulating accretion variability.

{Another useful comparison is with \citet{pittman25}, who analysed a large sample of discs and quantified their accretion regime by evaluating the ratio between the magnetospheric and co–rotation radii, expressed through $\omega_s=(R_M/R_{\rm co})^{3/2}$.  From the simulations of Ireland et al. (2022), systems with $\omega_s < 0.28$ are expected to accrete steadily, while those with $\omega_s \gtrsim 0.28$ should lie in the intermittent or propeller regimes. Overall, the fraction of steady versus intermittent/propeller systems in \citet{pittman25} is comparable to what we find in our smaller, but more homogeneous,  sample. Four sources in our sample are also included in the \citet{pittman25} analysis (DM Tau, J1842 and J1852 and AA Tau.). For these objects, the accretion state inferred from $\omega_s$ is broadly consistent with the position they occupy in our Figures \ref{fig:correlations} and \ref{fig:mdot_prot}. The only exception is J1852, for which the truncation radius estimate  of \citet{pittman25} is very small, thus locating it in the steady accretion regime. We however caution that empirical estimates of the truncation radius are uncertain, with the (kinematically determined)  values derived by \citet{pittman25}  being smaller than those inferred from interferometric measurements \citep{gravity23}.}

Finally we note that the large range of stellar rotation periods in our sample, which is otherwise rather homogeneous in terms of its stellar and current disc properties, is an expected outcome of braking models for pre-main sequence stars. This is partly because the rotation period reflects the prior history of disc braking and partly because the efficiency of braking also depends on the magnitude and time dependence of the stellar magnetic field, which can vary from star to star. Moreover the shortest period system, V4046 Sgr, is a tidally synchronised binary and it is to be expected that stars can also be spun up through tidal interaction with any planets orbiting within the corotation radius (see \citet{gallet18}, though note that these calculations did not include additional torques associated with star-disc interaction).

\subsubsection{Origin of viscosity}
The values of $\alpha$ derived in \citet{Longarini25} are independent of the specific physical mechanism responsible for angular momentum transport \citep{rosotti23}. Equation~\ref{alpha} quantifies the efficiency of transport that is required to reproduce the observed stellar accretion rate, whether this arises from turbulence, MHD processes, or magnetically--driven winds. By construction, this estimate assumes that the inner and outer disc are dynamically connected, so that the mass flux supplied at large radii ultimately reaches the star.  Within this framework, the correlation we find between accretion rate (and hence $\alpha$) and stellar rotation period is particularly significant. This trend does not depend on the detailed origin of viscosity in the outer disc, but rather reflects the regulation of mass transfer in the innermost regions through magnetospheric truncation and disc–star coupling. In other words, the $P_{\rm rot}$–$\dot{M}_\star$ correlation can be robustly established regardless of how angular momentum is transported on large scales.  

This reasoning also applies when considering wind--driven evolution. Although the wind scenario offers more flexibility than the viscous one---since the intial accretion timescale can vary from source to source \citep{tabone22}---this freedom is not sufficient to account for the large scatter in $\dot{M}_\star$ (and $\alpha$) without invoking temporal variability, since predicted disc lifetimes are otherwise far too long in some cases.  Moreover, the correlation between accretion rate and stellar rotation period strongly suggests that the origin of the variability lies in the innermost disc, where magnetospheric truncation regulates mass transfer, rather than in the details of angular momentum transport at large scales.

\subsection{Comments individual sources}
Several sources in our sample are known to exhibit variability, consistent with our interpretation of them being in an \textit{intermittent} or \textit{propeller} state.

We emphasise that the observational baseline for assessing accretion rate variations is very short
compared with the possible timescales on which a system evolves in the plane of rotation period 
versus accretion rate but it is nevertheless useful to summarise the available observational information
on accretion rate variability in individual sources. \citet{costigan14} report mean accretion-rate variations of $0.01$--$0.07$~dex on timescales shorter than one hour, $0.04$--$0.4$~dex over several days, and $0.13$--$0.52$~dex over years. These limits imply that while accretion can vary, short-term fluctuations are modest, whereas more substantial changes develop over longer  intervals.

{A well-documented example is LkCa15. The accretion rate adopted by \citet{Longarini25} to infer $\alpha$ was taken from \citet{manara14}, who measured $\dot{M}_\star \simeq 10^{-8.7}\,M_\odot\,{\rm yr^{-1}}$. A subsequent measurement by \citet{donati19} reported a lower value, $\dot{M}_\star \simeq 10^{-9.2}\,M_\odot\,{\rm yr^{-1}}$. We stress that a $\sim 0.5$~dex variation in $\dot M_\star$ is not, by itself, evidence for a change of accretion regime: such amplitudes are fully consistent with the short-term accretion variability commonly observed in T~Tauri stars, including the 2-week monitoring of LkCa15 by \citet{alencar18}. Both \citet{alencar18} and \citet{donati19} found that, within uncertainties, the truncation radius derived using the \citet{bessolaz08} prescription is compatible with the co-rotation radius.
Nevertheless, LkCa15 remains a compelling example in the context of our interpretation. Its stellar rotation period (5.7 days) lies extremely close to the $\sim 6$~day threshold at which systems in our sample show a sharp transition between high and low accretion rates. In this configuration—where $R_M$ and $R_{\rm co}$ are expected to be comparable—modest stochastic or rotationally-driven modulations can move the system between efficient and suppressed accretion, producing the type of variability seen in the observations without requiring large changes in $\dot M_\star$. Thus, while the observed accretion-rate excursions alone do not prove entry into the propeller regime, they are entirely consistent with a system hovering near the $R_M \approx R_{\rm co}$ boundary, where intermittent accretion is naturally expected.}


Finally, another interesting case is the close binary V4046 Sgr. \citet{stempels04} identified it as a double-lined spectroscopic binary with a 2.4-day orbital period, and argued that the rapid stellar rotation is maintained by tidal synchronisation within the binary. In our correlation plot this system lies at the extreme left, with one of the shortest rotation periods in the sample. Its anomalously low accretion rate may therefore not be due solely to magnetospheric regulation, but also to the combined effects of tidal interaction and star–disc coupling. {Recent high-resolution spectropolarimetric work by \citet{pouilly25} has further shown that the two components of V4046 Sgr accrete in qualitatively different regimes: the secondary appears to operate in a chaotic regime dominated by accretion tongues, the primary in a more ordered but still non-steady funnel/tongue hybrid (i.e., an “ordered-chaotic” regime). While this study does not explicitly frame these behaviours in terms of the classical propeller criterion (i.e. \(R_M>R_{\rm co}\)), the presence of unstable and non-steady accretion flows is entirely compatible with a scenario in which \(R_M\approx R_{\rm co}\) and the system sits near the boundary between steady and suppressed accretion. This indicates that a single-star magnetospheric accretion framework is insufficient to describe the system, and that binary dynamics must be taken into account.}

\section{Conclusions}\label{conclusions}
In this letter, we  investigated the wide spread of $\alpha$-viscosity values in a sample of well-characterised protoplanetary discs. Despite displaying similar global disc properties, such as masses and sizes, these systems show a broad range of accretion rates $\dot{M}_\star$ and hence of inferred $\alpha$ values.
We find that both $\dot{M}_\star$ and $\alpha$ correlate with the stellar rotation period. Rapidly rotating systems appear to be in a bottleneck state, where accretion is suppressed. This behaviour is consistent with magnetospheric truncation models \citep{ireland22}, in which the accretion regime is determined by the relative location of the magnetospheric radius $R_M$ and the co-rotation radius $R_{\rm co}$. For slowly rotating stars, the magnetosphere lies well inside co-rotation, and the outer disc efficiently channels material onto the star. Conversely, when the truncation radius approaches or exceeds the co-rotation radius (the propeller regime), accretion is strongly reduced and becomes intermittent, as angular momentum is injected into the disc material, inhibiting inflow.

Such a correlation between accretion rate and stellar rotation period has not been identified in earlier studies, most likely because the range of stellar and disc properties considered was much broader than in our carefully selected sample. We therefore suggest that the transition between steady accretion and the propeller state occurs at different stellar rotation periods depending on the specific disc properties. Further work is required to explore this dependence.

We thus propose that the observed spread in $\alpha$ reflects the transition between two distinct accretion regimes—a propeller state and a steady state—ultimately controlled by the stellar spin.

\section*{Acknowledgements}
{We thank the anonymous referee for the insightful comments and valuable feedback, which helped us to improve the paper.} The authors thank Jerome Bouvier, Doug Lin, Carlo Manara and Anik Halder for useful discussions. The authors have been supported by the UK Science and Technology Research Council (STFC) via the consolidated grant ST/W000997/1. CL acknowledges support from the COST Action CA22133 PLANETS.

\section*{Data Availability}
All the data underlying this article are publicly available in the ALMA Science Archive as part of the MAPS and exoALMA programmes. Stellar and disc parameters are taken from the literature (see references in the main text). Derived quantities and analysis scripts will be shared upon reasonable request to the corresponding author.

\bibliographystyle{mnras}
\bibliography{example} 

@ARTICLE{Longarini25,
       author = {{Longarini}, Cristiano and {Lodato}, Giuseppe and {Rosotti}, Giovanni and {Andrews}, Sean and {Winter}, Andrew and {Stadler}, Jochen and {Izquierdo}, Andr{\'e}s and {Galloway-Sprietsma}, Maria and {Facchini}, Stefano and {Curone}, Pietro and {Benisty}, Myriam and {Teague}, Richard and {Bae}, Jaehan and {Barraza-Alfaro}, Marcelo and {Cataldi}, Gianni and {Czekala}, Ian and {Cuello}, Nicol{\'a}s and {Fasano}, Daniele and {Flock}, Mario and {Fukagawa}, Misato and {Garg}, Himanshi and {Hall}, Cassandra and {Hammond}, Iain and {Hardiman}, Caitlyn and {Hilder}, Thomas and {Huang}, Jane and {Ilee}, John D. and {Isella}, Andrea and {Kanagawa}, Kazuhiro and {Lesur}, Geoffroy and {Loomis}, Ryan A. and {M{\'e}nard}, Francois and {Orihara}, Ryuta and {Pinte}, Christophe and {Price}, Daniel and {Testi}, Leonardo and {Fernandez}, Gaylor Wafflard- and {W{\"o}lfer}, Lisa and {Yen}, Hsi-Wei and {Yoshida}, Tomohiro C. and {Zawadzki}, Brianna},
        title = "{exoALMA. XII. Weighing and Sizing exoALMA Disks with Rotation Curve Modelling}",
      journal = {\apjl},
         year = 2025,
        month = may,
       volume = {984},
       number = {1},
          eid = {L17},
        pages = {L17},
          doi = {10.3847/2041-8213/adc431},
archivePrefix = {arXiv},
       eprint = {2504.18726},
 primaryClass = {astro-ph.EP},
       adsurl = {https://ui.adsabs.harvard.edu/abs/2025ApJ...984L..17L}
}

@ARTICLE{martire24,
       author = {{Martire}, P. and {Longarini}, C. and {Lodato}, G. and {Rosotti}, G.~P. and {Winter}, A. and {Facchini}, S. and {Hardiman}, C. and {Benisty}, M. and {Stadler}, J. and {Izquierdo}, A.~F. and {Testi}, Leonardo},
        title = "{Rotation curves in protoplanetary disks with thermal stratification. Physical model and observational evidence in MAPS disks}",
      journal = {\aap},
         year = 2024,
        month = jun,
       volume = {686},
          eid = {A9},
        pages = {A9},
          doi = {10.1051/0004-6361/202348546},
archivePrefix = {arXiv},
       eprint = {2402.12236},
 primaryClass = {astro-ph.EP},
       adsurl = {https://ui.adsabs.harvard.edu/abs/2024A&A...686A...9M}
}

@INPROCEEDINGS{hartmann88,
       author = {{Hartmann}, Lee and {Kenyon}, Scott},
        title = "{Accretion Disks Around Young Stars}",
    booktitle = {Formation and Evolution of Low Mass Stars},
         year = 1988,
       editor = {{Dupree}, A.~K. and {Lago}, M.~T.~V.~T.},
       series = {NATO Advanced Study Institute (ASI) Series C},
       volume = {241},
        month = jan,
        pages = {163},
          doi = {10.1007/978-94-009-3037-7_10},
       adsurl = {https://ui.adsabs.harvard.edu/abs/1988ASIC..241..163H}
}

@ARTICLE{artemenko12,
       author = {{Artemenko}, S.~A. and {Grankin}, K.~N. and {Petrov}, P.~P.},
        title = "{Rotation effects in classical T Tauri stars}",
      journal = {Astronomy Letters},
         year = 2012,
        month = dec,
       volume = {38},
       number = {12},
        pages = {783-792},
          doi = {10.1134/S1063773712110011},
archivePrefix = {arXiv},
       eprint = {1301.2493},
 primaryClass = {astro-ph.SR},
       adsurl = {https://ui.adsabs.harvard.edu/abs/2012AstL...38..783A}
}

@ARTICLE{percy2010,
       author = {{Percy}, John R. and {Grynko}, Sergiy and {Seneviratne}, Rajiv and {Herbst}, William},
        title = "{Self-Correlation Analysis of the Photometric Variability of T Tauri Stars. II. A Survey}",
      journal = {\pasp},
         year = 2010,
        month = jul,
       volume = {122},
       number = {893},
        pages = {753},
          doi = {10.1086/654826},
       adsurl = {https://ui.adsabs.harvard.edu/abs/2010PASP..122..753P}
}

@ARTICLE{siwack16,
       author = {{Siwak}, Michal and {Ogloza}, Waldemar and {Rucinski}, Slavek M. and {Moffat}, Anthony F.~J. and {Matthews}, Jaymie M. and {Cameron}, Chris and {Guenther}, David B. and {Kuschnig}, Rainer and {Rowe}, Jason F. and {Sasselov}, Dimitar and {Weiss}, Werner W.},
        title = "{Stable and unstable accretion in the classical T Tauri stars IM Lup and RU Lup as observed by MOST}",
      journal = {\mnras},
         year = 2016,
        month = mar,
       volume = {456},
       number = {4},
        pages = {3972-3984},
          doi = {10.1093/mnras/stv2848},
archivePrefix = {arXiv},
       eprint = {1512.01992},
 primaryClass = {astro-ph.SR},
       adsurl = {https://ui.adsabs.harvard.edu/abs/2016MNRAS.456.3972S}
}

@ARTICLE{stempels04,
       author = {{Stempels}, H.~C. and {Gahm}, G.~F.},
        title = "{The close T Tauri binary V 4046 Sagittarii}",
      journal = {\aap},
         year = 2004,
        month = jul,
       volume = {421},
        pages = {1159-1168},
          doi = {10.1051/0004-6361:20034502},
       adsurl = {https://ui.adsabs.harvard.edu/abs/2004A&A...421.1159S}
}

@ARTICLE{batalha98,
       author = {{Batalha}, C.~C. and {Quast}, G.~R. and {Torres}, C.~A.~O. and {Pereira}, P.~C.~R. and {Terra}, M.~A.~O. and {Jablonski}, F. and {Schiavon}, R.~P. and {de La Reza}, J.~R. and {Sartori}, M.~J.},
        title = "{Photometric variability of southern T Tauri stars}",
      journal = {\aaps},
         year = 1998,
        month = mar,
       volume = {128},
        pages = {561-571},
          doi = {10.1051/aas:1998163},
       adsurl = {https://ui.adsabs.harvard.edu/abs/1998A&AS..128..561B}
}

@ARTICLE{alencar18,
       author = {{Alencar}, S.~H.~P. and {Bouvier}, J. and {Donati}, J. -F. and {Alecian}, E. and {Folsom}, C.~P. and {Grankin}, K. and {Hussain}, G.~A.~J. and {Hill}, C. and {Cody}, A. -M. and {Carmona}, A. and {Dougados}, C. and {Gregory}, S.~G. and {Herczeg}, G. and {M{\'e}nard}, F. and {Moutou}, C. and {Malo}, L. and {Takami}, M. and {Matysse Collaboration}},
        title = "{Inner disk structure of the classical T Tauri star LkCa 15}",
      journal = {\aap},
         year = 2018,
        month = dec,
       volume = {620},
          eid = {A195},
        pages = {A195},
          doi = {10.1051/0004-6361/201834263},
archivePrefix = {arXiv},
       eprint = {1811.04806},
 primaryClass = {astro-ph.SR},
       adsurl = {https://ui.adsabs.harvard.edu/abs/2018A&A...620A.195A}
}

@ARTICLE{ireland22,
       author = {{Ireland}, Lewis G. and {Matt}, Sean P. and {Zanni}, Claudio},
        title = "{Magnetic Braking of Accreting T Tauri Stars II: Torque Formulation Spanning Spin-up and Spin-down Regimes}",
      journal = {\apj},
         year = 2022,
        month = apr,
       volume = {929},
       number = {1},
          eid = {65},
        pages = {65},
          doi = {10.3847/1538-4357/ac59b2},
archivePrefix = {arXiv},
       eprint = {2203.00326},
 primaryClass = {astro-ph.SR},
       adsurl = {https://ui.adsabs.harvard.edu/abs/2022ApJ...929...65I}
}

@ARTICLE{venuti17,
       author = {{Venuti}, L. and {Bouvier}, J. and {Cody}, A.~M. and {Stauffer}, J.~R. and {Micela}, G. and {Rebull}, L.~M. and {Alencar}, S.~H.~P. and {Sousa}, A.~P. and {Hillenbrand}, L.~A. and {Flaccomio}, E.},
        title = "{CSI 2264: Investigating rotation and its connection with disk accretion in the young open cluster NGC 2264}",
      journal = {\aap},
         year = 2017,
        month = mar,
       volume = {599},
          eid = {A23},
        pages = {A23},
          doi = {10.1051/0004-6361/201629537},
archivePrefix = {arXiv},
       eprint = {1610.08811},
 primaryClass = {astro-ph.SR},
       adsurl = {https://ui.adsabs.harvard.edu/abs/2017A&A...599A..23V}
}

@ARTICLE{gallet18,
       author = {{Gallet}, F. and {Bolmont}, E. and {Bouvier}, J. and {Mathis}, S. and {Charbonnel}, C.},
        title = "{Planetary tidal interactions and the rotational evolution of low-mass stars. The Pleiades' anomaly}",
      journal = {\aap},
         year = 2018,
        month = nov,
       volume = {619},
          eid = {A80},
        pages = {A80},
          doi = {10.1051/0004-6361/201833576},
archivePrefix = {arXiv},
       eprint = {1808.08728},
 primaryClass = {astro-ph.EP},
       adsurl = {https://ui.adsabs.harvard.edu/abs/2018A&A...619A..80G}
}

@ARTICLE{kratter16,
       author = {{Kratter}, Kaitlin and {Lodato}, Giuseppe},
        title = "{Gravitational Instabilities in Circumstellar Disks}",
      journal = {\araa},
         year = 2016,
        month = sep,
       volume = {54},
        pages = {271-311},
          doi = {10.1146/annurev-astro-081915-023307},
archivePrefix = {arXiv},
       eprint = {1603.01280},
 primaryClass = {astro-ph.SR},
       adsurl = {https://ui.adsabs.harvard.edu/abs/2016ARA&A..54..271K}
}

@ARTICLE{andrews09,
       author = {{Andrews}, Sean M. and {Wilner}, D.~J. and {Hughes}, A.~M. and {Qi}, Chunhua and {Dullemond}, C.~P.},
        title = "{Protoplanetary Disk Structures in Ophiuchus}",
      journal = {\apj},
         year = 2009,
        month = aug,
       volume = {700},
       number = {2},
        pages = {1502-1523},
          doi = {10.1088/0004-637X/700/2/1502},
archivePrefix = {arXiv},
       eprint = {0906.0730},
 primaryClass = {astro-ph.EP},
       adsurl = {https://ui.adsabs.harvard.edu/abs/2009ApJ...700.1502A}
}

@ARTICLE{andrews10,
       author = {{Andrews}, Sean M. and {Wilner}, D.~J. and {Hughes}, A.~M. and {Qi}, Chunhua and {Dullemond}, C.~P.},
        title = "{Protoplanetary Disk Structures in Ophiuchus. II. Extension to Fainter Sources}",
      journal = {\apj},
         year = 2010,
        month = nov,
       volume = {723},
       number = {2},
        pages = {1241-1254},
          doi = {10.1088/0004-637X/723/2/1241},
archivePrefix = {arXiv},
       eprint = {1007.5070},
 primaryClass = {astro-ph.SR},
       adsurl = {https://ui.adsabs.harvard.edu/abs/2010ApJ...723.1241A}
}

@ARTICLE{rafikov17,
       author = {{Rafikov}, Roman R.},
        title = "{Protoplanetary Disks as (Possibly) Viscous Disks}",
      journal = {\apj},
         year = 2017,
        month = mar,
       volume = {837},
       number = {2},
          eid = {163},
        pages = {163},
          doi = {10.3847/1538-4357/aa6249},
archivePrefix = {arXiv},
       eprint = {1701.02352},
 primaryClass = {astro-ph.EP},
       adsurl = {https://ui.adsabs.harvard.edu/abs/2017ApJ...837..163R}
}

@ARTICLE{ansdell18,
       author = {{Ansdell}, M. and {Williams}, J.~P. and {Trapman}, L. and {van Terwisga}, S.~E. and {Facchini}, S. and {Manara}, C.~F. and {van der Marel}, N. and {Miotello}, A. and {Tazzari}, M. and {Hogerheijde}, M. and {Guidi}, G. and {Testi}, L. and {van Dishoeck}, E.~F.},
        title = "{ALMA Survey of Lupus Protoplanetary Disks. II. Gas Disk Radii}",
      journal = {\apj},
         year = 2018,
        month = may,
       volume = {859},
       number = {1},
          eid = {21},
        pages = {21},
          doi = {10.3847/1538-4357/aab890},
archivePrefix = {arXiv},
       eprint = {1803.05923},
 primaryClass = {astro-ph.EP},
       adsurl = {https://ui.adsabs.harvard.edu/abs/2018ApJ...859...21A}
}

@ARTICLE{vandermarel21,
       author = {{van der Marel}, Nienke and {Birnstiel}, Til and {Garufi}, Antonio and {Ragusa}, Enrico and {Christiaens}, Valentin and {Price}, Daniel J. and {Sallum}, Steph and {Muley}, Dhruv and {Francis}, Logan and {Dong}, Ruobing},
        title = "{On the Diversity of Asymmetries in Gapped Protoplanetary Disks}",
      journal = {\aj},
         year = 2021,
        month = jan,
       volume = {161},
       number = {1},
          eid = {33},
        pages = {33},
          doi = {10.3847/1538-3881/abc3ba},
archivePrefix = {arXiv},
       eprint = {2010.10568},
 primaryClass = {astro-ph.EP},
       adsurl = {https://ui.adsabs.harvard.edu/abs/2021AJ....161...33V}
}

@ARTICLE{balbus91,
       author = {{Balbus}, Steven A. and {Hawley}, John F.},
        title = "{A Powerful Local Shear Instability in Weakly Magnetized Disks. I. Linear Analysis}",
      journal = {\apj},
         year = 1991,
        month = jul,
       volume = {376},
        pages = {214},
          doi = {10.1086/170270},
       adsurl = {https://ui.adsabs.harvard.edu/abs/1991ApJ...376..214B}
}

@ARTICLE{teague25,
       author = {{Teague}, Richard and {Benisty}, Myriam and {Facchini}, Stefano and {Fukagawa}, Misato and {Pinte}, Christophe and {Andrews}, Sean M. and {Bae}, Jaehan and {Barraza-Alfaro}, Marcelo and {Cataldi}, Gianni and {Cuello}, Nicol{\'a}s and {Curone}, Pietro and {Czekala}, Ian and {Fasano}, Daniele and {Flock}, Mario and {Galloway-Sprietsma}, Maria and {Garg}, Himanshi and {Hall}, Cassandra and {Hammond}, Iain and {Hilder}, Thomas and {Huang}, Jane and {Ilee}, John D. and {Izquierdo}, Andr{\'e}s F. and {Kanagawa}, Kazuhiro and {Lesur}, Geoffroy and {Lodato}, Giuseppe and {Longarini}, Cristiano and {Loomis}, Ryan A. and {Masset}, Fr{\'e}d{\'e}ric and {Menard}, Francois and {Orihara}, Ryuta and {Price}, Daniel J. and {Rosotti}, Giovanni and {Stadler}, Jochen and {Testi}, Leonardo and {Yen}, Hsi-Wei and {Wafflard-Fernandez}, Gaylor and {Wilner}, David J. and {Winter}, Andrew J. and {W{\"o}lfer}, Lisa and {Yoshida}, Tomohiro C. and {Zawadzki}, Brianna},
        title = "{exoALMA. I. Science Goals, Project Design, and Data Products}",
      journal = {\apjl},
         year = 2025,
        month = may,
       volume = {984},
       number = {1},
          eid = {L6},
        pages = {L6},
          doi = {10.3847/2041-8213/adc43b},
archivePrefix = {arXiv},
       eprint = {2504.18688},
 primaryClass = {astro-ph.EP},
       adsurl = {https://ui.adsabs.harvard.edu/abs/2025ApJ...984L...6T}
}

@ARTICLE{oberg21,
       author = {{{\"O}berg}, Karin I. and {Guzm{\'a}n}, Viviana V. and {Walsh}, Catherine and {Aikawa}, Yuri and {Bergin}, Edwin A. and {Law}, Charles J. and {Loomis}, Ryan A. and {Alarc{\'o}n}, Felipe and {Andrews}, Sean M. and {Bae}, Jaehan and {Bergner}, Jennifer B. and {Boehler}, Yann and {Booth}, Alice S. and {Bosman}, Arthur D. and {Calahan}, Jenny K. and {Cataldi}, Gianni and {Cleeves}, L. Ilsedore and {Czekala}, Ian and {Furuya}, Kenji and {Huang}, Jane and {Ilee}, John D. and {Kurtovic}, Nicolas T. and {Le Gal}, Romane and {Liu}, Yao and {Long}, Feng and {M{\'e}nard}, Fran{\c{c}}ois and {Nomura}, Hideko and {P{\'e}rez}, Laura M. and {Qi}, Chunhua and {Schwarz}, Kamber R. and {Sierra}, Anibal and {Teague}, Richard and {Tsukagoshi}, Takashi and {Yamato}, Yoshihide and {van't Hoff}, Merel L.~R. and {Waggoner}, Abygail R. and {Wilner}, David J. and {Zhang}, Ke},
        title = "{Molecules with ALMA at Planet-forming Scales (MAPS). I. Program Overview and Highlights}",
      journal = {\apjs},
         year = 2021,
        month = nov,
       volume = {257},
       number = {1},
          eid = {1},
        pages = {1},
          doi = {10.3847/1538-4365/ac1432},
archivePrefix = {arXiv},
       eprint = {2109.06268},
 primaryClass = {astro-ph.EP},
       adsurl = {https://ui.adsabs.harvard.edu/abs/2021ApJS..257....1O}
}

@ARTICLE{clarke95,
       author = {{Clarke}, C.~J. and {Armitage}, P.~J. and {Smith}, K.~W. and {Pringle}, J.~E.},
        title = "{Magnetically modulated accretion in T Tauri stars}",
      journal = {\mnras},
         year = 1995,
        month = apr,
       volume = {273},
       number = {3},
        pages = {639-642},
          doi = {10.1093/mnras/273.3.639},
archivePrefix = {arXiv},
       eprint = {astro-ph/9502061},
 primaryClass = {astro-ph},
       adsurl = {https://ui.adsabs.harvard.edu/abs/1995MNRAS.273..639C}
}

@ARTICLE{shakura73,
       author = {{Shakura}, N.~I. and {Sunyaev}, R.~A.},
        title = "{Black holes in binary systems. Observational appearance.}",
      journal = {\aap},
         year = 1973,
        month = jan,
       volume = {24},
        pages = {337-355},
       adsurl = {https://ui.adsabs.harvard.edu/abs/1973A&A....24..337S}
}

@ARTICLE{tabone22,
       author = {{Tabone}, Beno{\^\i}t and {Rosotti}, Giovanni P. and {Cridland}, Alexander J. and {Armitage}, Philip J. and {Lodato}, Giuseppe},
        title = "{Secular evolution of MHD wind-driven discs: analytical solutions in the expanded {\ensuremath{\alpha}}-framework}",
      journal = {\mnras},
         year = 2022,
        month = may,
       volume = {512},
       number = {2},
        pages = {2290-2309},
          doi = {10.1093/mnras/stab3442},
archivePrefix = {arXiv},
       eprint = {2111.10145},
 primaryClass = {astro-ph.SR},
       adsurl = {https://ui.adsabs.harvard.edu/abs/2022MNRAS.512.2290T}
}

@ARTICLE{law21b,
       author = {{Law}, Charles J. and {Teague}, Richard and {Loomis}, Ryan A. and {Bae}, Jaehan and {{\"O}berg}, Karin I. and {Czekala}, Ian and {Andrews}, Sean M. and {Aikawa}, Yuri and {Alarc{\'o}n}, Felipe and {Bergin}, Edwin A. and {Bergner}, Jennifer B. and {Booth}, Alice S. and {Bosman}, Arthur D. and {Calahan}, Jenny K. and {Cataldi}, Gianni and {Cleeves}, L. Ilsedore and {Furuya}, Kenji and {Guzm{\'a}n}, Viviana V. and {Huang}, Jane and {Ilee}, John D. and {Le Gal}, Romane and {Liu}, Yao and {Long}, Feng and {M{\'e}nard}, Fran{\c{c}}ois and {Nomura}, Hideko and {P{\'e}rez}, Laura M. and {Qi}, Chunhua and {Schwarz}, Kamber R. and {Soto}, Daniela and {Tsukagoshi}, Takashi and {Yamato}, Yoshihide and {van't Hoff}, Merel L.~R. and {Walsh}, Catherine and {Wilner}, David J. and {Zhang}, Ke},
        title = "{Molecules with ALMA at Planet-forming Scales (MAPS). IV. Emission Surfaces and Vertical Distribution of Molecules}",
      journal = {\apjs},
         year = 2021,
        month = nov,
       volume = {257},
       number = {1},
          eid = {4},
        pages = {4},
          doi = {10.3847/1538-4365/ac1439},
archivePrefix = {arXiv},
       eprint = {2109.06217},
 primaryClass = {astro-ph.GA},
       adsurl = {https://ui.adsabs.harvard.edu/abs/2021ApJS..257....4L}
}

@ARTICLE{galloway25,
       author = {{Galloway-Sprietsma}, Maria and {Bae}, Jaehan and {Izquierdo}, Andr{\'e}s F. and {Stadler}, Jochen and {Longarini}, Cristiano and {Teague}, Richard and {Andrews}, Sean M. and {Winter}, Andrew J. and {Benisty}, Myriam and {Facchini}, Stefano and {Rosotti}, Giovanni and {Zawadzki}, Brianna and {Pinte}, Christophe and {Fasano}, Daniele and {Barraza-Alfaro}, Marcelo and {Cataldi}, Gianni and {Cuello}, Nicol{\'a}s and {Curone}, Pietro and {Czekala}, Ian and {Flock}, Mario and {Fukagawa}, Misato and {Gardner}, Charles H. and {Garg}, Himanshi and {Hall}, Cassandra and {Huang}, Jane and {Ilee}, John D. and {Kanagawa}, Kazuhiro and {Lesur}, Geoffroy and {Lodato}, Giuseppe and {Loomis}, Ryan A. and {Menard}, Francois and {Orihara}, Ryuta and {Price}, Daniel J. and {Wafflard-Fernandez}, Gaylor and {Wilner}, David J. and {W{\"o}lfer}, Lisa and {Yen}, Hsi-Wei and {Yoshida}, Tomohiro C.},
        title = "{exoALMA. V. Gaseous Emission Surfaces and Temperature Structures}",
      journal = {\apjl},
         year = 2025,
        month = may,
       volume = {984},
       number = {1},
          eid = {L10},
        pages = {L10},
          doi = {10.3847/2041-8213/adc437},
archivePrefix = {arXiv},
       eprint = {2504.19902},
 primaryClass = {astro-ph.EP},
       adsurl = {https://ui.adsabs.harvard.edu/abs/2025ApJ...984L..10G}
}

@ARTICLE{manara14,
       author = {{Manara}, C.~F. and {Testi}, L. and {Natta}, A. and {Rosotti}, G. and {Benisty}, M. and {Ercolano}, B. and {Ricci}, L.},
        title = "{Gas content of transitional disks: a VLT/X-Shooter study of accretion and winds}",
      journal = {\aap},
         year = 2014,
        month = aug,
       volume = {568},
          eid = {A18},
        pages = {A18},
          doi = {10.1051/0004-6361/201323318},
archivePrefix = {arXiv},
       eprint = {1406.1428},
 primaryClass = {astro-ph.SR},
       adsurl = {https://ui.adsabs.harvard.edu/abs/2014A&A...568A..18M}
}

@ARTICLE{donati19,
       author = {{Donati}, J. -F. and {Bouvier}, J. and {Alencar}, S.~H. and {Hill}, C. and {Carmona}, A. and {Folsom}, C.~P. and {M{\'e}nard}, F. and {Gregory}, S.~G. and {Hussain}, G.~A. and {Grankin}, K. and {Moutou}, C. and {Malo}, L. and {Takami}, M. and {Herczeg}, G.~J. and {MaTYSSE Collaboration}},
        title = "{The magnetic propeller accretion regime of LkCa 15}",
      journal = {\mnras},
         year = 2019,
        month = feb,
       volume = {483},
       number = {1},
        pages = {L1-L5},
          doi = {10.1093/mnrasl/sly207},
archivePrefix = {arXiv},
       eprint = {1811.04810},
 primaryClass = {astro-ph.SR},
       adsurl = {https://ui.adsabs.harvard.edu/abs/2019MNRAS.483L...1D}
}

@ARTICLE{schaefer83,
       author = {{Schaefer}, B.~E.},
        title = "{A large amplitude photometric periodicity on a T Tauri star.}",
      journal = {\apjl},
         year = 1983,
        month = mar,
       volume = {266},
        pages = {L45-L49},
          doi = {10.1086/183975},
       adsurl = {https://ui.adsabs.harvard.edu/abs/1983ApJ...266L..45S}
}

@ARTICLE{rosotti23,
       author = {{Rosotti}, Giovanni P.},
        title = "{Empirical constraints on turbulence in proto-planetary discs}",
      journal = {\nar},
         year = 2023,
        month = jun,
       volume = {96},
          eid = {101674},
        pages = {101674},
          doi = {10.1016/j.newar.2023.101674},
archivePrefix = {arXiv},
       eprint = {2302.01433},
 primaryClass = {astro-ph.EP},
       adsurl = {https://ui.adsabs.harvard.edu/abs/2023NewAR..9601674R}
}

@ARTICLE{armitage96,
       author = {{Armitage}, P.~J. and {Clarke}, C.~J.},
        title = "{Magnetic braking of T Tauri stars}",
      journal = {\mnras},
         year = 1996,
        month = may,
       volume = {280},
       number = {2},
        pages = {458-468},
          doi = {10.1093/mnras/280.2.458},
archivePrefix = {arXiv},
       eprint = {astro-ph/9512018},
 primaryClass = {astro-ph},
       adsurl = {https://ui.adsabs.harvard.edu/abs/1996MNRAS.280..458A}
}

@ARTICLE{matt10,
       author = {{Matt}, Sean P. and {Pinz{\'o}n}, Giovanni and {de la Reza}, Ramiro and {Greene}, Thomas P.},
        title = "{Spin Evolution of Accreting Young Stars. I. Effect of Magnetic Star-Disk Coupling}",
      journal = {\apj},
         year = 2010,
        month = may,
       volume = {714},
       number = {2},
        pages = {989-1000},
          doi = {10.1088/0004-637X/714/2/989},
archivePrefix = {arXiv},
       eprint = {1005.0863},
 primaryClass = {astro-ph.SR},
       adsurl = {https://ui.adsabs.harvard.edu/abs/2010ApJ...714..989M}
}

@ARTICLE{matt12,
       author = {{Matt}, Sean P. and {Pinz{\'o}n}, Giovanni and {Greene}, Thomas P. and {Pudritz}, Ralph E.},
        title = "{Spin Evolution of Accreting Young Stars. II. Effect of Accretion-powered Stellar Winds}",
      journal = {\apj},
         year = 2012,
        month = jan,
       volume = {745},
       number = {1},
          eid = {101},
        pages = {101},
          doi = {10.1088/0004-637X/745/1/101},
archivePrefix = {arXiv},
       eprint = {1111.6407},
 primaryClass = {astro-ph.SR},
       adsurl = {https://ui.adsabs.harvard.edu/abs/2012ApJ...745..101M}
}

@ARTICLE{gallet19,
       author = {{Gallet}, F. and {Zanni}, C. and {Amard}, L.},
        title = "{Rotational evolution of solar-type protostars during the star-disk interaction phase}",
      journal = {\aap},
         year = 2019,
        month = dec,
       volume = {632},
          eid = {A6},
        pages = {A6},
          doi = {10.1051/0004-6361/201935432},
archivePrefix = {arXiv},
       eprint = {1910.03995},
 primaryClass = {astro-ph.SR},
       adsurl = {https://ui.adsabs.harvard.edu/abs/2019A&A...632A...6G}
}

@ARTICLE{pittman25,
       author = {{Pittman}, Caeley V. and {Espaillat}, C.~C. and {Zhu}, Zhaohuan and {Thanathibodee}, Thanawuth and {Robinson}, Connor E. and {Calvet}, Nuria and {K{\'o}sp{\'a}l}, {\'A}gnes},
        title = "{The ODYSSEUS Survey. Using accretion and stellar rotation to reveal the star-disk connection in T Tauri stars}",
      journal = {arXiv e-prints},
         year = 2025,
        month = sep,
          eid = {arXiv:2509.03767},
        pages = {arXiv:2509.03767},
          doi = {10.48550/arXiv.2509.03767},
archivePrefix = {arXiv},
       eprint = {2509.03767},
 primaryClass = {astro-ph.SR},
       adsurl = {https://ui.adsabs.harvard.edu/abs/2025arXiv250903767P}
}

@ARTICLE{deboer16,
       author = {{de Boer}, J. and {Salter}, G. and {Benisty}, M. and {Vigan}, A. and {Boccaletti}, A. and {Pinilla}, P. and {Ginski}, C. and {Juhasz}, A. and {Maire}, A. -L. and {Messina}, S. and {Desidera}, S. and {Cheetham}, A. and {Girard}, J.~H. and {Wahhaj}, Z. and {Langlois}, M. and {Bonnefoy}, M. and {Beuzit}, J. -L. and {Buenzli}, E. and {Chauvin}, G. and {Dominik}, C. and {Feldt}, M. and {Gratton}, R. and {Hagelberg}, J. and {Isella}, A. and {Janson}, M. and {Keller}, C.~U. and {Lagrange}, A. -M. and {Lannier}, J. and {Menard}, F. and {Mesa}, D. and {Mouillet}, D. and {Mugrauer}, M. and {Peretti}, S. and {Perrot}, C. and {Sissa}, E. and {Snik}, F. and {Vogt}, N. and {Zurlo}, A. and {SPHERE Consortium}},
        title = "{Multiple rings in the transition disk and companion candidates around RX J1615.3-3255. High contrast imaging with VLT/SPHERE}",
      journal = {\aap},
         year = 2016,
        month = nov,
       volume = {595},
          eid = {A114},
        pages = {A114},
          doi = {10.1051/0004-6361/201629267},
archivePrefix = {arXiv},
       eprint = {1610.04038},
 primaryClass = {astro-ph.EP},
       adsurl = {https://ui.adsabs.harvard.edu/abs/2016A&A...595A.114D}
}

@ARTICLE{romanova09,
       author = {{Romanova}, M.~M. and {Ustyugova}, G.~V. and {Koldoba}, A.~V. and {Lovelace}, R.~V.~E.},
        title = "{Launching of conical winds and axial jets from the disc-magnetosphere boundary: axisymmetric and 3D simulations}",
      journal = {\mnras},
         year = 2009,
        month = nov,
       volume = {399},
       number = {4},
        pages = {1802-1828},
          doi = {10.1111/j.1365-2966.2009.15413.x},
archivePrefix = {arXiv},
       eprint = {0907.3394},
 primaryClass = {astro-ph.SR},
       adsurl = {https://ui.adsabs.harvard.edu/abs/2009MNRAS.399.1802R}
}

@ARTICLE{pantolmos20,
       author = {{Pantolmos}, G. and {Zanni}, C. and {Bouvier}, J.},
        title = "{Magnetic torques on T Tauri stars: Accreting versus non-accreting systems}",
      journal = {\aap},
         year = 2020,
        month = nov,
       volume = {643},
          eid = {A129},
        pages = {A129},
          doi = {10.1051/0004-6361/202038569},
archivePrefix = {arXiv},
       eprint = {2009.00940},
 primaryClass = {astro-ph.SR},
       adsurl = {https://ui.adsabs.harvard.edu/abs/2020A&A...643A.129P}
}

@ARTICLE{costigan14,
       author = {{Costigan}, G. and {Vink}, Jorick S. and {Scholz}, A. and {Ray}, T. and {Testi}, L.},
        title = "{Temperaments of young stars: rapid mass accretion rate changes in T Tauri and Herbig Ae stars}",
      journal = {\mnras},
         year = 2014,
        month = jun,
       volume = {440},
       number = {4},
        pages = {3444-3461},
          doi = {10.1093/mnras/stu529},
archivePrefix = {arXiv},
       eprint = {1403.4088},
 primaryClass = {astro-ph.SR},
       adsurl = {https://ui.adsabs.harvard.edu/abs/2014MNRAS.440.3444C}
}

@ARTICLE{bessolaz08,
       author = {{Bessolaz}, N. and {Zanni}, C. and {Ferreira}, J. and {Keppens}, R. and {Bouvier}, J.},
        title = "{Accretion funnels onto weakly magnetized young stars}",
      journal = {\aap},
         year = 2008,
        month = jan,
       volume = {478},
       number = {1},
        pages = {155-162},
          doi = {10.1051/0004-6361:20078328},
archivePrefix = {arXiv},
       eprint = {0712.2921},
 primaryClass = {astro-ph},
       adsurl = {https://ui.adsabs.harvard.edu/abs/2008A&A...478..155B}
}

@ARTICLE{pouilly25,
       author = {{Pouilly}, K. and {Audard}, M.},
        title = "{Dissimilar magnetically driven accretion on the components of V4046 Sagittarii}",
      journal = {\aap},
         year = 2025,
        month = jun,
       volume = {698},
          eid = {A100},
        pages = {A100},
          doi = {10.1051/0004-6361/202554051},
archivePrefix = {arXiv},
       eprint = {2504.10217},
 primaryClass = {astro-ph.SR},
       adsurl = {https://ui.adsabs.harvard.edu/abs/2025A&A...698A.100P}
}

@ARTICLE{gravity23,
       author = {{Gravity Collaboration} and {Wojtczak}, J.~A. and {Labadie}, L. and {Perraut}, K. and {Tessore}, B. and {Soulain}, A. and {Ganci}, V. and {Bouvier}, J. and {Dougados}, C. and {Al{\'e}cian}, E. and {Nowacki}, H. and {Cozzo}, G. and {Brandner}, W. and {Caratti O Garatti}, A. and {Garcia}, P. and {Garcia Lopez}, R. and {Sanchez-Bermudez}, J. and {Amorim}, A. and {Benisty}, M. and {Berger}, J.-P. and {Bourdarot}, G. and {Caselli}, P. and {Cl{\'e}net}, Y. and {de Zeeuw}, P.~T. and {Davies}, R. and {Drescher}, A. and {Duvert}, G. and {Eckart}, A. and {Eisenhauer}, F. and {Eupen}, F. and {F{\"o}rster-Schreiber}, N.~M. and {Gendron}, E. and {Gillessen}, S. and {Grant}, S. and {Grellmann}, R. and {Hei{\ss}el}, G. and {Henning}, Th. and {Hippler}, S. and {Horrobin}, M. and {Hubert}, Z. and {Jocou}, L. and {Kervella}, P. and {Lacour}, S. and {Lapeyr{\`e}re}, V. and {Le Bouquin}, J.-B. and {L{\'e}na}, P. and {Lutz}, D. and {Mang}, F. and {Ott}, T. and {Paumard}, T. and {Perrin}, G. and {Scheithauer}, S. and {Shangguan}, J. and {Shimizu}, T. and {Spezzano}, S. and {Straub}, O. and {Straubmeier}, C. and {Sturm}, E. and {van Dishoeck}, E. and {Vincent}, F. and {Widmann}, F.},
        title = "{The GRAVITY young stellar object survey. IX. Spatially resolved kinematics of hot hydrogen gas in the star-disk interaction region of T Tauri stars}",
      journal = {\aap},
         year = 2023,
        month = jan,
       volume = {669},
          eid = {A59},
        pages = {A59},
          doi = {10.1051/0004-6361/202244675},
archivePrefix = {arXiv},
       eprint = {2210.13095},
 primaryClass = {astro-ph.SR},
       adsurl = {https://ui.adsabs.harvard.edu/abs/2023A&A...669A..59G}
}

@ARTICLE{Izquierdo25,
       author = {{Izquierdo}, Andr{\'e}s F. and {Stadler}, Jochen and {Galloway-Sprietsma}, Maria and {Benisty}, Myriam and {Pinte}, Christophe and {Bae}, Jaehan and {Teague}, Richard and {Facchini}, Stefano and {W{\"o}lfer}, Lisa and {Longarini}, Cristiano and {Curone}, Pietro and {Andrews}, Sean M. and {Barraza-Alfaro}, Marcelo and {Cataldi}, Gianni and {Cuello}, Nicol{\'a}s and {Czekala}, Ian and {Fasano}, Daniele and {Flock}, Mario and {Fukagawa}, Misato and {Garg}, Himanshi and {Hall}, Cassandra and {Hammond}, Iain and {Hilder}, Thomas and {Huang}, Jane and {Ilee}, John D. and {Isella}, Andrea and {Kanagawa}, Kazuhiro and {Lesur}, Geoffroy and {Lodato}, Giuseppe and {Loomis}, Ryan A. and {Orihara}, Ryuta and {Price}, Daniel J. and {Rosotti}, Giovanni and {Testi}, Leonardo and {Yen}, Hsi-Wei and {Wafflard-Fernandez}, Gaylor and {Wilner}, David J. and {Winter}, Andrew J. and {Yoshida}, Tomohiro C. and {Zawadzki}, Brianna},
        title = "{exoALMA. III. Line-intensity Modeling and System Property Extraction from Protoplanetary Disks}",
      journal = {\apjl},
         year = 2025,
        month = may,
       volume = {984},
       number = {1},
          eid = {L8},
        pages = {L8},
          doi = {10.3847/2041-8213/adc439},
archivePrefix = {arXiv},
       eprint = {2504.19986},
 primaryClass = {astro-ph.EP},
       adsurl = {https://ui.adsabs.harvard.edu/abs/2025ApJ...984L...8I}
}
\bsp	
\label{lastpage}
\end{document}